# Spin-dependent transport in $p^+$-$CdB_xF_{2-x}$ – $n$-$CdF_2$ planar structures

**N T Bagraev[1], M I Bovt[2], O N Guimbitskaya[2], L E Klyachkin[1], A M Malyarenko[1], A I Ryskin[3] and A S Shcheulin[3]**

[1] Ioffe Physico-Technical Institute RAS, 194021, St.Petersburg, Russia
[2] St.Petersburg Polytechnical University, St.Petersburg, 195251, Russia
[3] Vavilov State Optical Institute, St.Petersburg, 199034, Russia

E-mail: impurity.dipole@mail.ioffe.ru

**Abstract**. The CV measurements and tunneling spectroscopy are used to study the ballistic transport of the spin-polarized holes by varying the value of the Rashba spin-orbit interaction (SOI) in the p-type quantum well prepared on the surface of the $n$-$CdF_2$<Y> bulk crystal. The findings of the hole conductance oscillations in the plane of the p-type quantum well that are due to the variations of the Rashba SOI are shown to be evidence of the spin transistor effect, with the amplitude of the oscillations close to $e^2/h$.

The ionic semiconductor crystal $CdF_2$ is of extraordinary interest for the modern optics and optoelectronics because of the largest band-gap value, 7.8 eV, from all wide-gap semiconductors [1, 2]. This dielectric as-grown crystal has been well documented to be converted into n-type semiconductor by doping with the donor impurities that represent the III group elements and subsequent thermal annealing in the reduction atmosphere with a colouration revealed during this process [1]. The monopolar origin of the $CdF_2$ crystals that exhibit only the n-type conductivity is however a serious obstacle for the application in semiconductor electronics. Nevertheless, boron atoms appear to be the best candidates to replace the positions of interstitial fluorine atoms in the $CdF_2$ lattice thereby converting the conductivity from the n-type to p-type by forming thin layers $CdB_xF_{2-x}$ on the surface of the bulk $CdF_2$ crystal [3]. Besides, the fluctuations in the distribution of boron inside these layers are able to give rise to the self-organization of the p-type $CdF_2$ nanostructures embedded in the $CdB_2$ shells that seem to be high temperature superconductors. The 2D holes inside such nanostructures are expected to be spin-polarized as a result of the sp-f exchange interaction through the rare-earth ions located in the $CdF_2$ bulk crystals. Furthermore, the availability of the $p^+$-n junction makes it possible to control the value of the Rashba spin-orbit interaction (SOI) in the p-type quantum well by varying the forward and reverse bias voltage [4]. Both these factors are attributable to the spin transistor effect caused by the spin precession due to the Rashba SOI that results in the current modulation [5]. Therefore the goal of this work is to apply the short-time diffusion of boron into the n-type $CdF_2$ bulk crystal heavily doped with yttrium, 0.15%, for the preparation of the p-type high mobility $CdB_xF_{2-x}$ quantum well to demonstrate the spin transistor effect by controlling the Rashba (SOI) value.

The $p^+$-$CdB_xF_{2-x}$ – $n$-$CdF_2$ planar structures have been prepared in frameworks of the Hall geometry by the short-time diffusion of boron from the gas phase of the temperature of $620^0$C (figures 1 and 2). The low temperature gold-related technology has been developed to deposit the ohmic contacts to the $p^+$- $CdB_xF_{2-x}$ layers and to the top gate. The specific contacts to the n-type $CdF_2$ bulk crystal have been prepared by sputtering the ytterbium silizides.

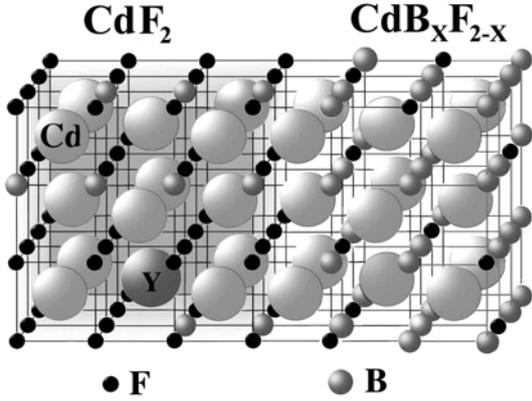

**Figure 1.** The $p^+$-$CdB_xF_{2-x}$ - $n$-$CdF_2$ junction that contains one-dimensional wires of boron inserted in the $CdF_2$ bulk crystal.

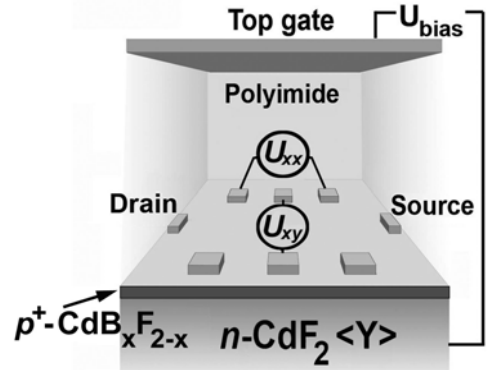

**Figure 2.** The scheme of the p-type $CdB_xF_{2-x}$ quantum well prepared on the n-$CdF_2$ surface, which is provided by the top gate to vary the value of the Rashba spin-orbit interaction.

The forward branches of the CV characteristics reveal not only the $CdF_2$ gap value, 7.8 eV, but also the $CdF_2$ valence band thoroughly as well thereby identifying the high temperature ballistic transport of holes in the $p^+$-n junction (figure 3a). The density of states in the $CdF_2$ valence band emerged from the CV fine structure is important to be in a good agreement with the data obtained using the photoemission technique [2]. Noteworthy also is the high value of the forward current that seems to be caused by the formation of the quasi-one-dimensional p-$CdB_2$ nanostructures embedded in the n-$CdF_2$ bulk crystal. In frameworks of such one-dimensional $p^+$-n junctions, the chains of boron atoms, which replace the fluorine atoms in the process of diffusion, pass smoothly in the $F^-$ chains (figure 1). Besides, the low energy part of the forward CV characteristic exhibits the spectrum of the tunnelling current that is related to the energy positions of the two-dimensional hole subbands in the p-type ultra-narrow quantum well prepared on the n-type $CdF_2$ surface (figures 3b and 4a). In addition to the tunneling current spectrum, the mesoscopic fluctuations of the conductance, $2e^2/h$, are observed, which are evidence of the hole quantum subbands caused by the Andreev reflection in the barriers confining the p-type $CdB_xF_{2-x}$ quantum well (figures 3c and 4b). These findings confirm the hypothesis proposed above that the p-type $CdB_xF_{2-x}$ nanostructures are self-assembled inside superconductor shells on the n-type $CdF_2$ surface. Moreover, the THz generation, 0.5 THz – 13 THz, that is attended with the mesoscopic conductance fluctuations appears to result in the high mobility of the 2D hole gas which is revealed by the Hall measurements: $8.3 \times 10^4$ cm$^2$/Vs at T = 300K ($p_{2D}$ = $2.5 \times 10^{12}$ cm$^{-2}$); and $96 \times 10^4$ cm$^2$/Vs at T = 77K ($p_{2D}$ = $2.6 \times 10^{11}$ cm$^{-2}$).

The studies of the static magnetic susceptibility have shown that the 2D holes in the p-type $CdB_xF_{2-x}$ quantum well are spin-polarized by the sp-f exchange interaction through the $Y^{3+}$ ions located in the n-$CdF_2$ bulk crystal. Thus, the properties of the $p^+$-$CdB_xF_{2-x}$ – n-$CdF_2$ nanostructures offer the prospect of the high temperature spin transistor effect without the injection of the spin-polarized carriers from the iron contacts as proposed in Ref [5]. This version of the spin transistor effect appears to be revealed by the conductance oscillations of the 2D holes that are found in the device studied (figure 2) by varying the top gate voltage, which controls the Rashba SOI value (figures 5a and 5b). The symmetrical behaviour of the conductance oscillations with the amplitude close to $e^2/h$ at different signs of the top gate voltage as well as their quenching by increasing the drain-source voltage support our contention (figures 5c and 5d).

The modulation of the conductance shown in figures 5a and 5b appears to result from the phase shift induced by the effective magnetic field, which is created by the Rashba SOI and represents the vector product of the external electric field and the carrier wavevector [6,7],

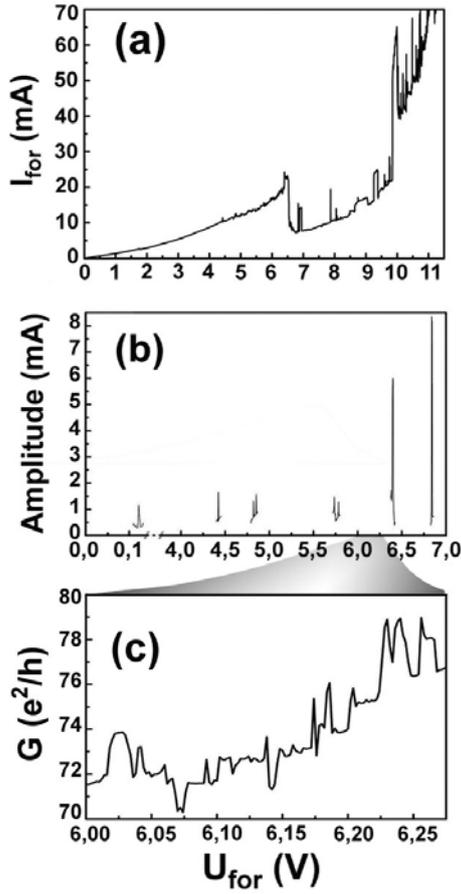
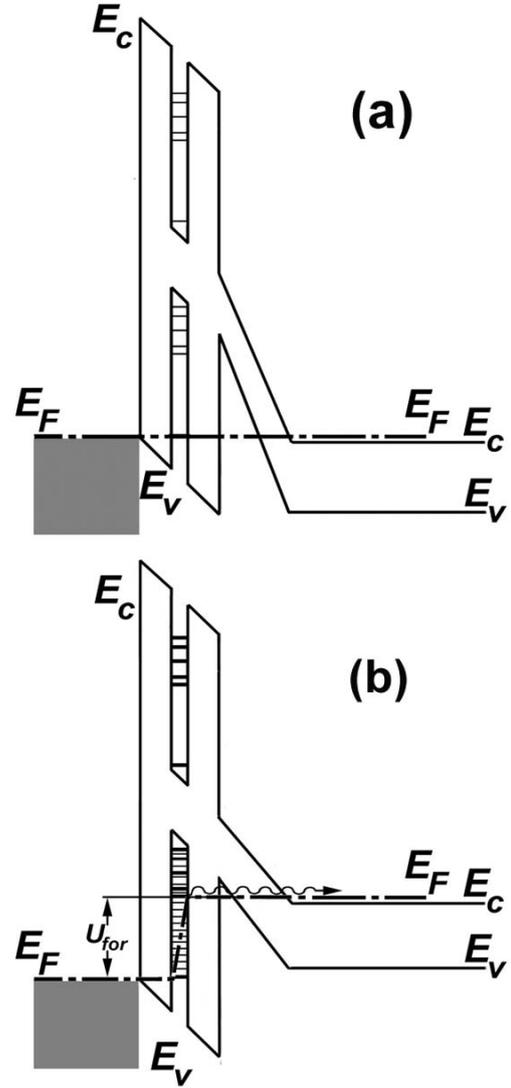

**Figure 3.** (a) The forward current-voltage characteristic revealed by the p+-$CdB_xF_{2-x}$ - n-$CdF_2$ junction that demonstrates the gap value, 7.8 eV, and the $CdF_2$ valence band structure.
(b) The tunneling part of the forward current-voltage characteristic that identifies the energy positions of the two-dimensional subbands of holes in the p-type $CdB_xF_{2-x}$ quantum well.
(c) Mesoscopic fluctuations of the conductance, $2e^2/h$, caused by the Andreev reflection in the barriers confining the p-type $CdB_xF_{2-x}$ quantum well, which define the spectrum of the THz generation: 0.5 THz - 13 THz. T = 300 K

**Figure 4.** (a), (b) The one-electron band scheme of the p-type $CdB_xF_{2-x}$ quantum well on the n-$CdF_2$ surface that demonstrates the two-dimensional subbands of holes (a) and the hole quantum subbands caused by the Andreev reflection in the barriers confining the p-type $CdB_xF_{2-x}$ quantum well (b).

$$\mathbf{B}_{eff} = \frac{\alpha}{g_B \mu_B} [\mathbf{k} \times \mathbf{e}_z] \qquad (1)$$

where $\mu_B$ is the Bohr magneton and

$$\alpha = -3\beta_{hh} \langle k_r^2 \rangle E_z \qquad (2)$$

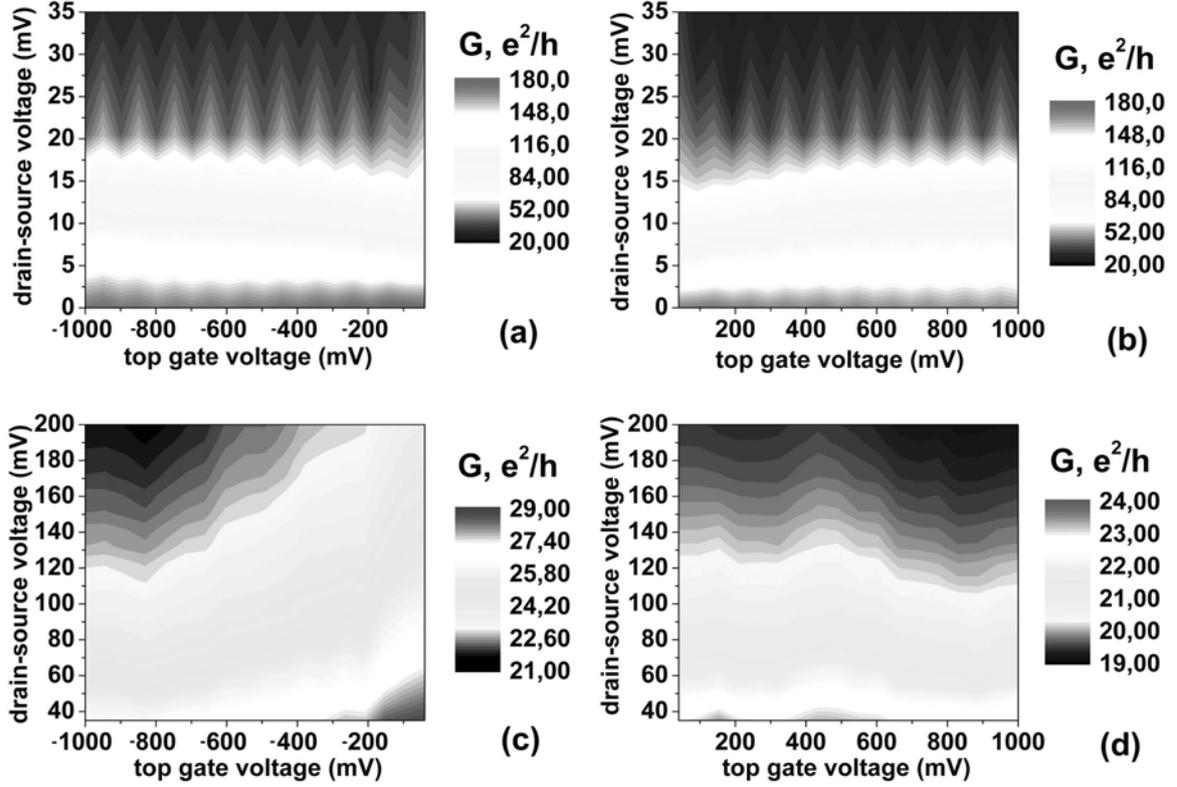

**Figure 5.** The conductance oscillations observed in the studies of the p-type $CdB_xF_{2-x}$ quantum well on the n-$CdF_2$ bulk crystal doped with yttrium that exhibit the spin transistor effect at T=300 K by varying the top gate voltage, which controls the value of the Rashba spin-orbit interaction ((a) and (b)). The drain-source current was stabilized, $I_{ds}$ = 10 nA. These conductance oscillations appear to smooth over increasing the drain-source voltage that gives rise to both the drop of the spin-lattice relaxation time and the transitions between different subbands of holes ((c) and (d)).

is the Rashba parameter; $E_z$ is determined by the value of the top gate voltage, $V_g = E_z l$, where $l$ is a characteristic length, which gives the proportionality between the gate voltage $V_g$ and the electric field applied perpendicularly to the quantum well interface; $\langle k_r^2 \rangle$ and $\beta_{hh}$ are dependent on the width of the quantum well and the energy positions of the 2D hole subbands. These relationships define the period of the conductance oscillations found as the principal characteristic of the spin transistor effect, which is determined by the characteristics of the device [6,7,8],

$$\Delta V_g \approx \frac{\hbar^2 d^2 l}{3\pi^2 R m_{eff} \beta_{hh}} \qquad (3)$$

where $d$ is the width of the quantum well, $R$ is the transport length. Thus, the experimental data presented in figures 5a and 5b allow the estimations of the value of the hole effective mass, if the energy positions of the subbands of the 2D holes obtained by the measurements of the CV tunneling characteristics are taken into account (figure 3b). This estimation results in extremely low effective mass of the 2D holes in the p-type $CdB_xF_{2-x}$ quantum well, 3.44 x $10^{-4}$ $m_0$, that seems to substantiate the observation of the spin transistor effect at T=300K.

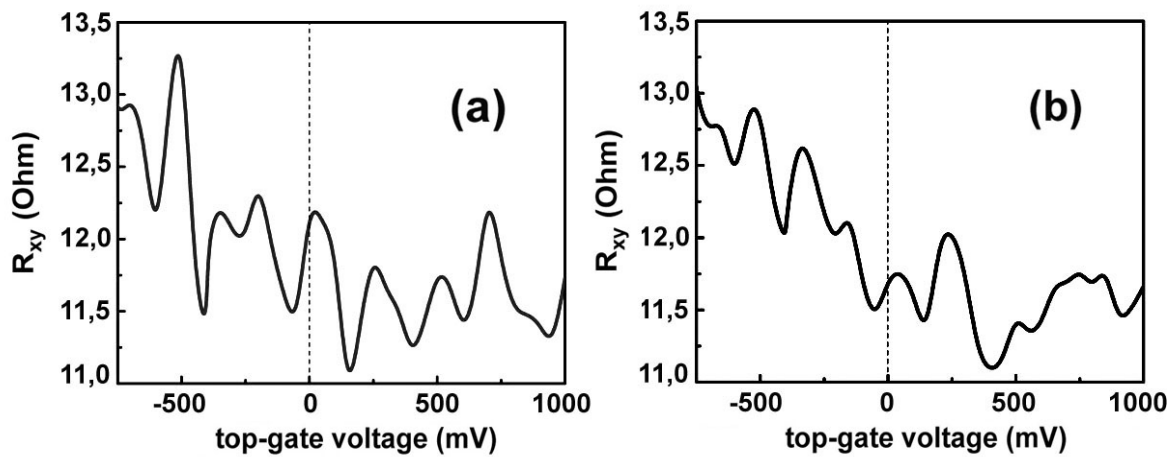

**Figure 6.** The variations of the Hall voltage caused by the Rashba spin-orbit interaction that is controlled by varying the top gate voltage applied to the p-type $CdB_xF_{2-x}$ quantum well on the n-$CdF_2$ bulk crystal doped with yttrium. (a)- $U_{ds}$=3 µV, (b) - 5 µV.

Finally, the effective magnetic field, $B_{eff}$, that is in parallel to $U_{xy}$ in the quantum well plane appears to give rise to the creation of the Hall voltage (figures 6a and 6b), because the spin-polarized 2D holes with a single spin direction, spin-up or spin-down, dominate the device studied as opposed to the classical version of the spin transistor that proposed the absence of the spin Hall effect [5].

In conclusions, the ultra-shallow $p^+$-$CdB_xF_{2-x}$ - n-$CdF_2$ structures have been prepared by doping with boron from the gas phase on the surface of the $CdF_2$(Y) bulk crystal, which exhibits the n-type conductivity as a result of previous thermal colouring. The forward CV characteristics of the $p^+$- n junctions prepared have been revealed both the $CdF_2$ valence band in detail and the high mobility p-type ultra-narrow quantum well on the n-type $CdF_2$ surface. The Hall measurements and the studies of static magnetic susceptibility have demonstrated the spin polarization of the 2D holes that is caused by the sp-f exchange interaction through the $Y^{3+}$ ions. The spin transistor and spin Hall effects have been found by varying the Rashba SOI value.

This work was supported by SNSF in frameworks of the programme "Scientific Cooperation between Eastern Europe and Switzerland, Grant IB7320-110970/1.